\documentclass[10pt,conference]{IEEEtran}

\usepackage{amsmath}
\usepackage{amssymb}
\usepackage[dvips]{graphicx}
\usepackage{setspace}

\renewcommand{\baselinestretch}{0.98}

\newcommand{\rr}{{\bf r}}
\newcommand{\bb}{{\bf b}}
\newcommand{\nn}{{\bf n}}

\newcommand{\xx}{{\bf x}}
\newcommand{\yy}{{\bf y}}
\newcommand{\zz}{{\bf z}}
\newcommand{\ee}{{\bf e}}
\newcommand{\qq}{{\bf q}}

\newcommand{\one}{{\bf 1}}
\newcommand{\zero}{{\bf 0}}

\newcommand{\ssT}{{\bf s}^T}
\newcommand{\ssV}{{\bf s}}

\newcommand{\xxx}{{\bf X}}
\newcommand{\rrr}{{\bf R}}
\newcommand{\sss}{{\bf S}}
\newcommand{\aaa}{{\bf A}}
\newcommand{\iii}{{\bf I}}

\newcommand{\ppp}{{\bf P}}

\newcommand{\mtL}{{\mathcal{L}}}

\newcommand{\MU}{\boldsymbol{\mu}}
\newcommand{\NU}{\boldsymbol{\nu}}

\begin{document}

\title{Optimal and Suboptimal Finger Selection Algorithms for MMSE Rake Receivers in Impulse Radio Ultra-Wideband Systems$^\textrm{\small{1}}$}
\author{\authorblockN{\normalsize Sinan Gezici, Mung Chiang, H. Vincent Poor and Hisashi Kobayashi}
\authorblockA{\normalsize Department of Electrical
Engineering\\\normalsize Princeton University, Princeton, NJ
08544\\\normalsize\{sgezici,chiangm,poor,hisashi\}@princeton.edu}}

\maketitle

\footnotetext[1]{This research is supported in part by the
National Science Foundation under grants ANI-03-38807,
CNS-0417603, and CCR-0440443, and in part by the New Jersey Center
for Wireless Telecommunications.}

\maketitle

\begin{abstract}
Convex relaxations of the optimal finger selection algorithm are
proposed for a minimum mean square error (MMSE) Rake receiver in
an impulse radio ultra-wideband system. First, the optimal finger
selection problem is formulated as an integer programming problem
with a non-convex objective function. Then, the objective function
is approximated by a convex function and the integer programming
problem is solved by means of constraint relaxation techniques.
The proposed algorithms are suboptimal due to the approximate
objective function and the constraint relaxation steps. However,
they can be used in conjunction with the conventional finger
selection algorithm, which is suboptimal on its own since it
ignores the correlation between multipath components, to obtain
performances reasonably close to that of the optimal scheme that
cannot be implemented in practice due to its complexity. The
proposed algorithms leverage convexity of the optimization problem
formulations, which is the watershed between `easy' and
`difficult' optimization problems.

\end{abstract}

\section{Introduction}

Since the US Federal Communications Commission (FCC) approved the
limited use of ultra-wideband (UWB) technology \cite{FCC},
communications systems that employ UWB signals have drawn
considerable attention. A UWB signal is defined to be one that
possesses an absolute bandwidth larger than $500$MHz or a relative
bandwidth larger than 20\% and can coexist with incumbent systems
in the same frequency range due to its large spreading factor and
low power spectral density. UWB technology holds great promise for
a variety of applications such as short-range high-speed data
transmission and precise location estimation.

Commonly, impulse radio (IR) systems, which transmit very short
pulses with a low duty cycle, are employed to implement UWB
systems (\cite{scholtz}-\cite{andy}). In an IR system, a train of
pulses is sent and information is usually conveyed by the position
or the polarity of the pulses, which correspond to Pulse Position
Modulation (PPM) and Binary Phase Shift Keying (BPSK),
respectively. In order to prevent catastrophic collisions among
different users and thus provide robustness against
multiple-access interference, each information symbol is
represented by a sequence of pulses; the positions of the pulses
within that sequence are determined by a pseudo-random
time-hopping (TH) sequence specific to each user \cite{scholtz}.
The number $N_f$ of pulses representing one information symbol can
also be interpreted as pulse combining gain.

Commonly, users in an IR-UWB system employ Rake receivers to
collect energy from different multipath components. A Rake
receiver combining all the paths of the incoming signal is called
an \textit{all-Rake} (\textit{ARake}) receiver. Since a UWB signal
has a very wide bandwidth, the number of resolvable multipath
components is usually very large. Hence, an ARake receiver is not
implemented in practice due to its complexity. However, it serves
as a benchmark for the performance of more practical Rake
receivers. A feasible implementation of multipath diversity
combining can be obtained by a \textit{selective-Rake}
(\textit{SRake}) receiver, which combines the $M$ best, out of
$L$, multipath components \cite{cassiICC02}. Those $M$ best
components are determined by a finger selection algorithm. For a
maximal ratio combining (MRC) Rake receiver, the paths with
highest signal-to-noise ratios (SNRs) are selected, which is an
optimal scheme in the absence of interfering users and
inter-symbol interference (ISI). For a minimum mean square error
(MMSE) Rake receiver, the ``conventional" finger selection
algorithm is to choose the paths with highest
signal-to-interference-plus-noise ratios (SINRs).
This conventional scheme is not necessarily optimal since it
ignores the correlation of the noise terms at different multipath
components. In other words, choosing the paths with highest SINRs
does not necessarily maximizes the overall SINR of the system.

In this paper, we formulate the optimal MMSE SRake as a nonconvex,
integer-constrained optimization, where the aim is to choose the
finger locations of the receiver so as to maximize the overall
SINR. While computing the optimal finger selection is NP-hard, we
present several relaxation methods to turn the (approximate)
problem into convex optimization problems that can be very
efficiently solved by interior-point methods, which are polynomial
time in the worst case, and very fast in practice. These optimal
finger selection relaxations produce significantly higher average
SINR than the conventional one that ignores the correlations, and
represent a numerically efficient way to strike a balance between
SINR optimality and computational tractability.

\newcounter{sec}
\setcounter{sec}{2} The remainder of the paper is organized as
follows. Section \Roman{sec} describes the transmitted and
received signal models in a multiuser frequency-selective
environment. The finger selection problem is formulated and the
optimal algorithm is described in Section
\setcounter{sec}{3}\Roman{sec}, which is followed by a brief
description of the conventional algorithm in Section
\setcounter{sec}{4}\Roman{sec}. In Section
\setcounter{sec}{5}\Roman{sec}, two convex relaxations of the
optimal finger selection algorithm, based on an approximate SINR
expression and integer constraint relaxation techniques, are
proposed. The simulation results are presented in Section
\setcounter{sec}{6}\Roman{sec}, and the concluding remarks are
made in the last section.

\section{Signal Model}

\begin{figure}
\begin{center}
\includegraphics[width=0.5\textwidth]{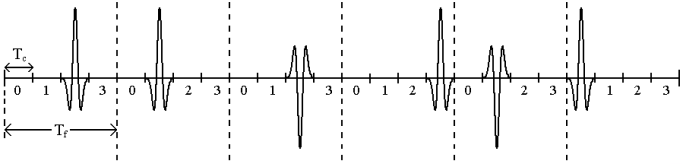}
\caption{An example time-hopping impulse radio signal with
pulse-based polarity randomization, where $N_f=6$, $N_c=4$, the
time hopping sequence is \{2,1,2,3,1,0\} and the polarity codes
are \{+1,+1,-1,+1,-1,+1\}.} \label{fig:TH_IR}
\end{center}
\end{figure}

We consider a synchronous, binary phase shift keyed TH-IR system
with $K$ users, in which the transmitted signal from user $k$ is
represented by:
\begin{gather}\label{eq:tran1}
s^{(k)}_{tx}(t)=\sqrt{\frac{E_k}{N_f}}\sum_{j=-\infty}^{\infty}d^{(k)}_j\,
b^{(k)}_{\lfloor j/N_f\rfloor}p_{tx}(t-jT_f-c^{(k)}_jT_c),
\end{gather}
where $p_{tx}(t)$ is the transmitted UWB pulse, $E_k$ is the bit
energy of user $k$, $T_f$ is the ``frame" time, $N_f$ is the number of pulses
representing one information symbol, and $b^{(k)}_{\lfloor j/N_f
\rfloor}\in \{+1,-1\}$ is the binary information symbol
transmitted by user $k$. In order to allow the channel to be
shared by many users and avoid catastrophic collisions, a
time-hopping (TH) sequence $\{c^{(k)}_j\}$, where $c^{(k)}_j \in
\{0,1,...,N_c-1\}$, is assigned to each user. This TH sequence
provides an additional time shift of $c^{(k)}_jT_c$ seconds to the
$j$th pulse of the $k$th user where $T_c$ is the chip interval
and is chosen to satisfy $T_c\leq T_f/N_c$ in order to prevent the
pulses from overlapping. We assume $T_f=N_cT_c$ without loss of generality.
The random polarity codes $d^{(k)}_j$ are binary random variables
taking values $\pm1$ with equal probability (\cite{eran1}-\cite{paul}).

Consider the discrete presentation of the channel,
$\boldsymbol{\alpha}^{(k)}=[\alpha_1^{(k)}\cdots \alpha_L^{(k)}]$
for user $k$, where $L$ is assumed to be the number of multipath
components for each user, and $T_c$ is the multipath resolution.
Then, the received signal can be expressed as
\begin{align}\label{eq:recSig}\nonumber
r(t)=\sum_{k=1}^{K}&\sqrt{\frac{E_k}{N_f}}\sum_{j=-\infty}^{\infty}\sum_{l=1}^{L}\alpha_l^{(k)}d^{(k)}_j\,
b^{(k)}_{\lfloor j/N_f\rfloor}\\
&\times p_{rx}(t-jT_f-c^{(k)}_jT_c-(l-1)T_c)+\sigma_nn(t),
\end{align}
where $p_{rx}(t)$ is the received unit-energy UWB pulse, which is
usually modelled as the derivative of $p_{tx}(t)$ due to the
effects of the antenna, and $n(t)$ is zero mean white Gaussian
noise with unit spectral density.

We assume that the time-hopping sequence is constrained to the set
$\{0,1,\ldots,N_T-1\}$, where $N_T\leq N_c-L$, so that there is no
inter-frame interference (IFI).

\begin{figure}
\begin{center}
\includegraphics[width=0.5\textwidth]{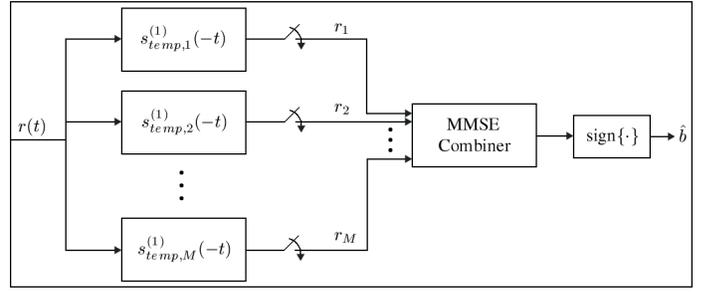}
\caption{The receiver structure. There are $M$ multipath
components, which are combined by the MMSE combiner.}
\label{fig:rec}
\end{center}
\end{figure}

Due to the high resolution of UWB signals, chip-rate and frame
rate sampling are not very practical for such systems. In order to
have a lower sampling rate, the received signal can be correlated
with template signals which enable symbol rate sampling of the
output \cite{molischWPMC03}. The template signal for the $l$th
path of the incoming signal can be expressed as
\begin{gather}\label{eq:temp}
s_{temp,l}^{(1)}(t)=\sum_{j=iN_f}^{(i+1)N_f-1}d^{(1)}_j\,
p_{rx}(t-jT_f-c^{(1)}_jT_c-(l-1)T_c),
\end{gather}
for the $i$th information symbol, where we consider user $1$
without loss of generality. In other words, by using a correlator
for each multipath component that we want to combine, we can just
use symbol rate sampling at each branch, as shown in Figure
\ref{fig:rec}.

Note that the use of such template signals results in equal gain
combining (EGC) of different frame components. This may not be
optimal under some conditions (see \cite{sinanUWBST04} for
(sub)optimal schemes). However, it is very practical since it
facilitates symbol-rate sampling. Since we consider a system that
employs template signals of the form (\ref{eq:temp}), i.e. EGC of
frame components, it is sufficient to consider the problem of
selection of the optimal paths just for one frame. Hence, we
assume $N_f=1$ without loss of generality.

Let $\mtL=\{l_1,\ldots,l_M\}$ denote the set of multipath
components that the receiver collects (Figure \ref{fig:rec}). At
each branch, the signal is effectively passed through a matched
filter (MF) matched to the related template signal in
(\ref{eq:temp}) and sampled once for each symbol. Then, the
discrete signal for the $l$th path can be expressed, for the $i$th
information symbol, as$^{2}$\footnotetext[2]{Note that the
dependence of $r_{l}$ on the index of the information symbol, $i$,
is not shown explicitly.}
\begin{gather}\label{eq:r_lj}
r_{l}=\ssT_{l}\aaa\bb_i+n_{l},
\end{gather}
for $l=l_1,\ldots,l_M$, where
$\aaa=\textrm{diag}\{\sqrt{E_1},\ldots,\sqrt{E_K}\}$,
$\bb_i=[b_i^{(1)}\cdots b_i^{(K)}]^T$ and
$n_{l}\sim\mathcal{N}(0\,,\,\sigma_n^2)$. $\ssV_{l}$ is a
$K\times1$ vector, which can be expressed as a sum of the desired
signal part (SP) and multiple-access interference (MAI) terms:
\begin{gather}\label{eq:sig_Vec}
\ssV_{l}=\ssV_{l}^{(SP)}+\ssV_{l}^{(MAI)},
\end{gather}
where the $k$th elements can be expressed as
\begin{align}\label{eq:sig_Vec_SP}
\left[\ssV_{l}^{(SP)}\right]_k&=
\begin{cases}
\alpha_l^{(1)},\quad\quad k=1\\
0,\quad\quad\quad\,k=2,\ldots,K
\end{cases}\textrm{and}\\
\left[\ssV_{l}^{(MAI)}\right]_k&=
\begin{cases}\label{eq:sig_Vec_MAI}
0,\quad\quad\quad\quad\quad\quad\quad\quad\quad\quad k=1\\
d_1^{(1)}d_1^{(k)}\sum_{m=1}^{L}\alpha_m^{(k)}I_{l,m}^{(k)},\,\,\,\,k=2,\ldots,K
\end{cases},
\end{align}
with $I_{l,m}^{(k)}$ being the indicator function that is equal to
$1$ if the $m$th path of user $k$ collides with the $l$th path of
user $1$, and $0$ otherwise.

\section{Problem Formulation and Optimal Solution}

The problem is to choose the optimal set of multipath components,
$\mathcal{L}=\{l_1,\ldots,l_M\}$, that minimizes the bit error
probability (BEP) of the system. In other words, we need to choose
the best samples from the $L$ received samples $r_l$,
$l=1,\ldots,L$, as shown in (\ref{eq:r_lj}).

To reformulate this combinatorial problem, we first define an
$M\times L$ selection matrix $\xxx$ as follows: $M$ of the columns
of $\xxx$ are the unit vectors $\ee_1,\ldots,\ee_M$ ($\ee_i$
having a $1$ at its $i$th position and zero elements for all other
entries), and the other columns are all zero vectors. The column
indices of the unit vectors determine the subset of the multipath
components that are selected. For example, for $L=4$ and $M=2$,
$\xxx=\begin{bmatrix}0\,\,1\,\,0\,\,0\\0\,\,0\,\,1\,\,0\end{bmatrix}$
chooses the second and third multipath components.

Using the selection matrix $\xxx$, we can express the vector of
received samples from $M$ multipath components as
\begin{gather}\label{eq:r1}
\rr=\xxx\sss\aaa\bb_i+\xxx\nn,
\end{gather}
where $\nn$ is the vector of thermal noise components
$\nn=[n_1\cdots n_L]^T$, and $\sss$ is the signature matrix given
by $\sss=[\ssV_1\cdots\ssV_{L}]^T$, with $\ssV_l$ as in
(\ref{eq:sig_Vec}).

Using (\ref{eq:sig_Vec})-(\ref{eq:sig_Vec_MAI}), (\ref{eq:r1}) can
be expressed as
\begin{gather}
\rr=b_i^{(1)}\sqrt{E_1}\xxx\boldsymbol{\alpha}^{(1)}+\xxx\sss^{(MAI)}\aaa\bb_i+\xxx\nn,
\end{gather}
where $\sss^{(MAI)}$ is the MAI part of the signature matrix
$\sss$.

Then, the linear MMSE receiver can be expressed as
\begin{gather}
\hat{b}_i=\textrm{sign}\{\boldsymbol{\theta}^T\rr\},
\end{gather}
where the MMSE weight vector is given by \cite{verdu}
\begin{gather}
\boldsymbol{\theta}=\rrr^{-1}\xxx\boldsymbol{\alpha}^{(1)},
\end{gather}
with $\rrr$ being the correlation matrix of the noise term:
\begin{gather}
\rrr=\xxx\sss^{(MAI)}\aaa^2(\sss^{(MAI)})^T\xxx^T+\sigma_n^2\iii.
\end{gather}

The SINR of the system can be expressed as
\begin{align}\label{eq:SINR}\nonumber
&SINR(\xxx)=\frac{E_1}{\sigma_n^2}(\boldsymbol{\alpha}^{(1)})^T\xxx^T
\\&\times\left(\iii+\frac{1}{\sigma_n^2}\xxx\sss^{(MAI)}\aaa^2(\sss^{(MAI)})^T\xxx^T\right)^{-1}\xxx\boldsymbol{\alpha}^{(1)}.
\end{align}
Hence, the optimal path/finger selection problem can be formulated
as
\begin{gather}\label{eq:optProb}
\textrm{maximize}\,\,SINR(\xxx),
\end{gather}
where $\xxx$ has the previously defined structure.

Note that the objective function to be maximized is not concave
and the optimization variable $\xxx$ takes binary values, with the
previously defined structure. In other words, two major
difficulties arise in solving (\ref{eq:optProb}) globally:
nonconvex optimization and integer constraints. Either makes the
problem NP-hard. Therefore, it is an intractable optimization
problem in this general form.

\section{Conventional Algorithm}

Instead of the solving the problem in (\ref{eq:optProb}), the
``conventional" finger selection algorithm chooses the $M$ paths
with largest individual SINRs, where the SINR for the $l$th path
can be expressed as
\begin{gather}\label{eq:indivSINR}
SINR_l=\frac{E_1(\alpha_l^{(1)})^2}{(\ssV_l^{(MAI)})^T\aaa^2\ssV_l^{(MAI)}+\sigma_n^2},
\end{gather}
for $l=1,\ldots,L$.

This algorithm is not optimal because it ignores the correlation
of the noise components of different paths. Therefore, it does not
always maximize the overall SINR of the system given in
(\ref{eq:SINR}). For example, the contribution of two highly
correlated strong paths to the overall SINR might be worse than
the contribution of one strong and one relatively weaker, but
uncorrelated, paths. The correlation between the multipath
components is the result of the MAI from the other users in the
system.

\section{Relaxations of Optimal Finger Selection}\label{sec:Subopt}

Since the optimal solution in (\ref{eq:optProb}) is quite
difficult, we first consider an approximation of the objective
function in (\ref{eq:SINR}). When the eigenvalues of
$\frac{1}{\sigma_n^2}\xxx\sss^{(MAI)}\aaa^2(\sss^{(MAI)})^T\xxx^T$
are considerably smaller than $1$, which occurs when the MAI is
not very strong compared to the thermal noise, we can approximate
the SINR expression in (\ref{eq:SINR}) as
follows$^{3}$\footnotetext[3]{More accurate approximations can be
obtained by using higher order series expansions for the matrix
inverse in (\ref{eq:SINR}). However, the solution of the
optimization problem does not lend itself to low complexity
solutions in those cases.}:
\begin{align}\label{eq:SINR2}\nonumber
&SINR(\xxx)\approx\frac{E_1}{\sigma_n^2}(\boldsymbol{\alpha}^{(1)})^T\xxx^T
\\&\times\left(\iii-\frac{1}{\sigma_n^2}\xxx\sss^{(MAI)}\aaa^2(\sss^{(MAI)})^T\xxx^T\right)\xxx\boldsymbol{\alpha}^{(1)},
\end{align}
which can be expressed as
\begin{align}\label{eq:SINR3}\nonumber
S&INR(\xxx)\approx\frac{E_1}{\sigma_n^2}\{\,(\boldsymbol{\alpha}^{(1)})^T\xxx^T\xxx\boldsymbol{\alpha}^{(1)}
\\&-\frac{1}{\sigma_n^2}\boldsymbol{\alpha}^{(1)}\xxx^T\xxx\sss^{(MAI)}\aaa^2(\sss^{(MAI)})^T\xxx^T\xxx\boldsymbol{\alpha}^{(1)}\,\}.
\end{align}
Note that the approximate $SINR$ expression depends on $\xxx$ only
through $\xxx^T\xxx$. Defining $\xx=[x_1\cdots x_L]^T$ as the
diagonal elements of $\xxx^T\xxx$,
$\xx=\textrm{diag}\{\xxx^T\xxx\}$, we have $x_i=1$ if the $i$th
path is selected, and $x_i=0$ otherwise; and
$\sum_{i=1}^{L}x_i=M$. Then, we obtain, after some manipulation,
\begin{gather}
SINR(\xx)=\frac{E_1}{\sigma_n^2}\left\{\qq^T\xx-\frac{1}{\sigma_n^2}\xx^T\ppp\xx\right\},
\end{gather}
where $\qq=[(\alpha_1^{(1)})^2\cdots(\alpha_L^{(1)})^2]^T$ and
$\ppp=\textrm{diag}\{\alpha_1^{(1)}\cdots\alpha_L^{(1)}\}\sss^{(MAI)}\aaa^2(\sss^{(MAI)})^T\textrm{diag}\{\alpha_1^{(1)}\cdots\alpha_L^{(1)}\}$.

Then, we can formulate the finger selection problem as follows:
\begin{align}\label{eq:subopt}\nonumber
\textrm{minimize}\,\quad&\frac{1}{\sigma_n^2}\xx^T\ppp\xx-\xx^T\qq\\\nonumber
\textrm{subject to}\quad&\xx^T\one=M,\\
&x_i\in\{0,1\},\quad i=1,\ldots,L.
\end{align}

Note that the objective function is convex since $\ppp$ is
positive definite, and that the first constraint is linear.
However, the integer constraint increases the complexity of the
problem. The common way to approximate the solution of an integer
constraint problem is to use \textit{constraint relaxation}. Then,
the optimizer will be a continuous value instead of being binary
and the problem (\ref{eq:subopt}) will be convex.
Over the past decade, both powerful theory and efficient numerical
algorithms have been developed for nonlinear convex optimization.
It is now recognized that the watershed between ``easy" and
``difficult" optimization problems is not linearity but convexity.
For example, the interior-point algorithms for nonlinear convex
optimization are highly efficient, both in worst case complexity
(provably polynomial time) and in practice (very fast even for a
large number variables and constraints) \cite{optTextbook}.
Interior-point methods solve convex optimization problems with
inequality constraints by applying Newton's method to a sequence
of equality constrained problems, where the Newton's method is a
kind of descent algorithm with the descent direction given by the
Newton step \cite{optTextbook}.

We consider two different relaxation techniques in the following
subsections.

\subsection{Case-1: Relaxation to Sphere}\label{sec:sphere}

Consider the relaxation of the integer constraint in
(\ref{eq:subopt}) to a sphere that passes through all possible
integer values. Then, the relaxed problem becomes
\begin{align}\label{eq:subopt2}\nonumber
\textrm{minimize}\,\quad&\frac{1}{\sigma_n^2}\xx^T\ppp\xx-\xx^T\qq\\\nonumber
\textrm{subject to}\quad&\,\xx^T\one=M,\\
&(2\xx-\one)^T(2\xx-\one)\leq L.
\end{align}
Note that the problem becomes a convex quadratically constrained
quadratic programming (QCQP) \cite{optTextbook}. Hence it can be
solved for global optimality using interior-point algorithms in
polynomial time.

\subsection{Case-2: Relaxation to Hypercube}\label{sec:hypercube}

As an alternative approach, we can relax the integer constraint in
(\ref{eq:subopt}) to a hypercube constraint and get
\begin{align}\label{eq:subopt3}\nonumber
\textrm{minimize}\,\quad&\frac{1}{\sigma_n^2}\xx^T\ppp\xx-\xx^T\qq\\\nonumber
\textrm{subject to}\quad&\,\xx^T\one=M,\\
&\,\xx\in[0,1]^L,
\end{align}
where the hypercube constraint can be expressed as
$\xx\succeq\bold{0}$ and $\xx\preceq\one$, with $\yy\succeq\zz$
meaning that $y_1\geq z_1,\ldots,y_L\geq z_L$. Note that the
problem is now a linearly constrained quadratic programming
(LCQP), and can be solved by interior-point algorithms
\cite{optTextbook} for the optimizer $\xx^{*}$.

\subsection{Dual Methods}

We can also consider the dual problems. For the relaxation to the
sphere considered in Section \ref{sec:sphere}, the Lagrangian for
(\ref{eq:subopt2}) can be obtained as
\begin{gather}
\mathcal{L}(\xx,\lambda,\nu)=\xx^T\left(\frac{1}{\sigma_n^2}\ppp+4\nu\iii\right)\xx-\xx^T(\qq-\lambda\one+4\nu\one)-M\lambda,
\end{gather}
where $\lambda\in\mathcal{R}$ and $\nu\in\mathcal{R}^{+}$.

After some manipulation, the Lagrange dual function can be
expressed as
\begin{align}\nonumber
g(\lambda,\nu)&=-\frac{1}{4}\,[\qq+(\lambda+4\nu)\one]^T\left(\frac{1}{\sigma_n^2}\ppp+4\nu\iii\right)^{-1}
\\&[\qq+(\lambda+4\nu)\one]-M\lambda,
\end{align}
Then, the dual problem becomes
\begin{align}\label{eq:subopt2_dual}\nonumber
&\textrm{minimize}\\&\frac{1}{4}[\qq+(\lambda+4\nu)\one]^T\left(\frac{1}{\sigma_n^2}\ppp+4\nu\iii\right)^{-1}[\qq+(\lambda+4\nu)\one]+M\lambda\\
&\textrm{subject to}\quad\nu\geq 0,
\end{align}
which can be solved for optimal $\lambda$ and $\nu$ by interior
point methods. Or, more simply, the unconstraint problem
(\ref{eq:subopt2_dual}) can be solved using gradient descent
algorithm, and then the optimizer $\bar{\nu}$ is mapped to
$\nu^{*}=\max\{0,\bar{\nu}\}$.

After solving for optimal $\lambda$ and $\mu$, the optimizer
$\xx^{*}$ is obtained as
\begin{gather}
\xx^{*}=\frac{1}{2}\left(\frac{1}{\sigma_n^2}\ppp+4\nu^{*}\iii\right)^{-1}[\qq+(\lambda^{*}+4\nu^{*})\one].
\end{gather}

Note that the dual problem (\ref{eq:subopt2_dual}) has two
variables, $\lambda$ and $\nu$, to optimize, compared to $L$
variables, the components of $\xx$, in the primal problem
(\ref{eq:subopt2}). However, an $L\times L$ matrix needs to be
inverted for each iteration of the optimization of
(\ref{eq:subopt2_dual}). Therefore, the primal problem can be
preferred over the dual problem in this case.

Similarly, the dual problem for the relaxation in Section
\ref{sec:hypercube} can be obtained from (\ref{eq:subopt3}) as
\begin{align}\label{eq:subopt3_dual}\nonumber
&\textrm{minimize}\\
&\frac{\sigma_n^2}{4}(\qq+\MU-\NU-\lambda\one)^T\ppp^{-1}(\qq+\MU-\NU-\lambda\one)+M\lambda+\NU^T\one\\
&\textrm{subject to}\quad\MU,\NU\succeq \zero.
\end{align}

It is observed from (\ref{eq:subopt3_dual}) that there are $2L+1$
variables and also $L\times L$ matrix inversion operations for the
solution of the dual problem. Therefore, the simpler primal
problem (\ref{eq:subopt3}) is considered in the simulations.

\subsection{Selection of Finger Locations}

After solving the approximate problem (\ref{eq:subopt}) by means
of integer relaxation techniques mentioned above, the finger
location estimations are obtained by the indices of the $M$
largest elements of the optimizer $\xx^{*}$.

Both the approximation of the SINR expression by (\ref{eq:SINR2})
and the integer relaxation steps result in the suboptimality of
the solution. Therefore, it may not be very close to the optimal
solution in some cases. However, it is expected to perform better
than the conventional algorithm most of the time, since it
considers the correlation between the multipath components.
However, it is not guaranteed that the algorithms based on the
convex relaxations of optimal finger selection always beat the
conventional one. Since the conventional algorithm is very easy to
implement, we can consider a hybrid algorithm where the final
estimate of the convex relaxation algorithm is compared with that
of the conventional one and the one that minimizes the exact SINR
expression in (\ref{eq:SINR}) is chosen as the final estimate. In
this way, the resulting hybrid suboptimal algorithm can get closer
to the optimal solution.

\section{Simulation Results}

The simulation results are performed to evaluate the performance
of different finger selection algorithms for an IR-UWB system with
$N_c=20$ and $N_f=1$. There are $5$ equal energy users in the
system ($K=5$) and the users' TH and polarity codes are randomly
generated. We model the channel coefficients as
$\alpha_l=\textrm{sign}(\alpha_l)|\alpha_l|$ for $l=1,\ldots,L$,
where $\textrm{sign}(\alpha_l)$ is $\pm1$ with equal probability
and $|\alpha_l|$ is distributed lognormally as
$\mathcal{LN}(\mu_l,\sigma^2)$. Also the energy of the taps is
exponentially decaying as
$\textrm{E}\{|\alpha_l|^2\}=\Omega_0e^{-\lambda(l-1)}$, where
$\lambda$ is the decay factor and
$\sum_{l=1}^{L}\textrm{E}\{|\alpha_l|^2\}=1$ (so
$\Omega_0=(1-e^{-\lambda})/(1-e^{-\lambda L})$). For the channel
parameters, we have $\lambda=0.1$, $\sigma^2=0.5$ and $\mu_l$ can
be calculated from
$\mu_l=0.5\left[\textrm{ln}(\frac{1-e^{-\lambda}}{1-e^{-\lambda
L}})-\lambda(l-1)-2\sigma^2\right]$, for $l=1,\ldots,L$. We
average the overall SINR of the system over different realizations
of channel coefficients, TH and polarity codes of the users.

\begin{figure}
\begin{center}
\includegraphics[width=0.5\textwidth]{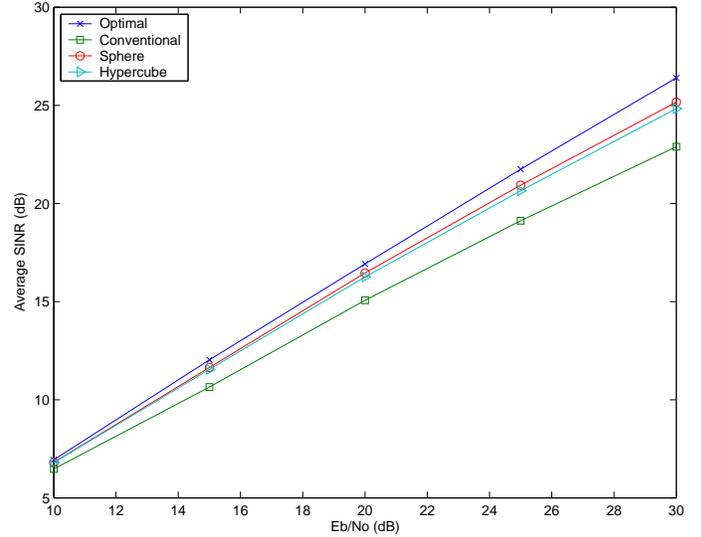}
\caption{\small Average SINR versus $E_b/N_0$ for $M=5$ fingers.
The channel has $L=15$ multipath components and the taps are
exponentially decaying. The IR-UWB system has $N_c=20$ chips per
frame and $N_f=1$ frame per symbol. There are $5$ equal energy
users in the system and random TH and polarity codes are used.}
\label{fig:SINR_vs_EbNo}
\end{center}
\end{figure}

In Figure \ref{fig:SINR_vs_EbNo}, we plot the average SINR of the
system for different noise variances when $M=5$ fingers are to be
chosen out of $L=15$ multipath components. As is observed from the
figure, the convex relaxations of optimal finger selection result
in SINR values reasonable close to those of the optimal exhaustive
search scheme. Note that the gain by using the proposed algorithms
over the conventional one increases as the thermal noise
decreases. This is because when the thermal noise get less
significant, the MAI becomes dominant, and the conventional
technique gets worse since it ignores the correlation between the
MAI noise terms when choosing the fingers.

\begin{figure}
\begin{center}
\includegraphics[width=0.5\textwidth]{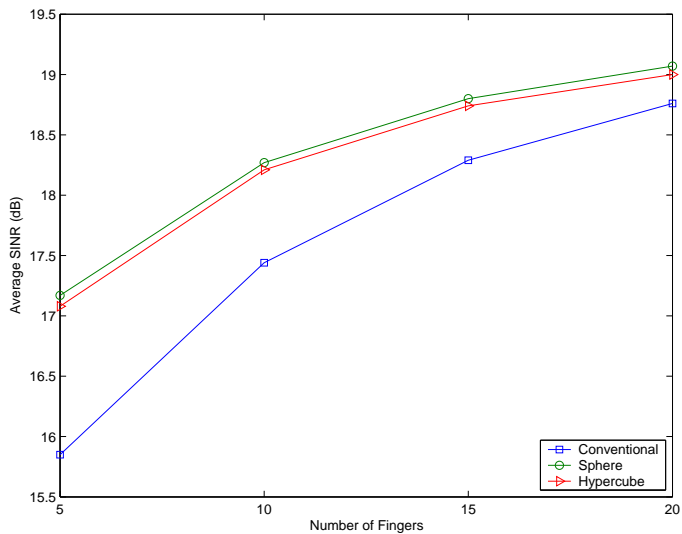}
\caption{\small Average SINR versus number of fingers $M$, for
$E_b/N_0=20$dB, $N_c=75$ and $L=50$. All the other parameters are
the same as those for Figure \ref{fig:SINR_vs_EbNo}.}
\label{fig:SINR_vs_M_2}
\end{center}
\end{figure}

Next, we plot SINR of the proposed suboptimal and conventional
techniques for different finger numbers in Figure
\ref{fig:SINR_vs_M_2}, where there are $50$ multipath components
and $E_b/N_0=20$. The number of chips per frame, $N_c$, is set to
$75$, and all other parameters are kept the same. In this case,
the optimal algorithm takes a very long time to simulate since it
needs to perform exhaustive search over many different finger
combinations (therefore not implemented). The improvement using
convex relaxations of optimal finger selection over the
conventional technique decreases as $M$ gets large since the
channel is exponentially decaying and the most of the significant
multipath components are already combined by all the algorithms.

\begin{figure}
\begin{center}
\includegraphics[width=0.5\textwidth]{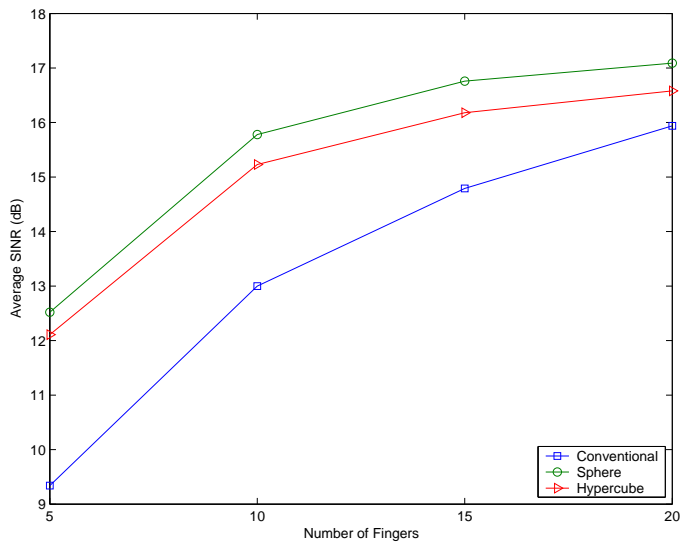}
\caption{\small Average SINR versus number of fingers $M$. There
are $10$ users with each interferer having $10$dB more power than
the desired user. All the other parameters are the same as those
for Figure \ref{fig:SINR_vs_M_2}.} \label{fig:SINR_vs_M_MAIlim}
\end{center}
\end{figure}

Finally, we consider a MAI-limited scenario, where there are $10$
users with $E_1=1$ and $E_k=10$ $\forall k$, and all the
parameters are as in the previous case. Then, as shown in Figure
\ref{fig:SINR_vs_M_MAIlim}, the improvement by using the
suboptimal finger selection algorithms increase significantly. The
main reason for this is that the suboptimal algorithms consider,
although approximately, the correlation caused by MAI whereas the
conventional scheme simply ignores that.

\section{Concluding Remarks}

Optimal and suboptimal finger selection algorithms for MMSE-SRake
receivers in an IR-UWB system are considered. Since UWB systems
have large number of multipath components, only a subset of those
components can be used due to complexity constraints. Therefore,
the selection of the optimal subset of multipath components is
important for the performance of the receiver. We have shown that
the optimal solution to this finger selection problem requires
exhaustive search which becomes prohibitive for UWB systems.
Therefore, we have proposed approximate solutions of the problem
based on the Taylor series approximation and integer constraint
relaxations. Using two different integer relaxation approaches, we
have introduced two convex relaxations of the optimal finger
selection algorithm. Implementing these suboptimal algorithms on
top of the conventional scheme, we can get close to the optimal
solution, with much lower complexity.

The two contributions of the paper are the formulation of the
optimal problem and the convex relaxations. In the first, the
formulation is globally optimal but the solution methods for
non-convex nonlinear integer constrained optimization must use
heuristics to get to a locally optimal solution because otherwise
computational load for global optimality is too much. In the
second, the formulation is relaxed, but the interior-point methods
efficiently computes the globally optimal solution for these
relaxations.

\end{document}